\documentclass[aps,prl,reprint,amsmath,amssymb,superscriptaddress]{revtex4-2}

\usepackage[T1]{fontenc}
\usepackage{lmodern}
\usepackage{bm}
\usepackage{graphicx}
\usepackage{microtype}
\usepackage{tikz}
\usetikzlibrary{arrows.meta,positioning}
\usepackage[hidelinks]{hyperref}

\newcommand{\Cov}{\operatorname{Cov}}

\begin{document}

\title{Stationarity as a One-Mode Constraint on Quantum Correlation}

\author{Itai Panas}
\affiliation{Department of Chemistry and Chemical Engineering, Chalmers University of Technology, SE-412 96 Gothenburg, Sweden}

\date{October 12, 2026}

\begin{abstract}
A stationary scalar correlation energy admits an exact one-mode terminal representation even when the microscopic correlated return is strongly multimode. We test predictive realizations built from specified stationary information carriers in two complementary two-electron systems with the same Coulomb repulsion but different confinement: He-like ions and harmonium (Hooke's atom). For the Coulomb sequence, canonical RHF supplies the symmetry-faithful information carrier. The repeated same-mode return gives a He correlation error of $+0.1164$ m$E_{\mathrm h}$ and no more than $0.0157$ m$E_{\mathrm h}$ for Li$^+$--Ar$^{16+}$; H$^-$ is nevertheless bound and $93.3\%$ of its conventional correlation energy is recovered even though the RHF carrier is qualitatively strained. The dimensionless confinement factor $R_{CC}=D_{CC}/Z$ tends to $1.092394$. For harmonic confinement, an independently constructed response-compressed one-mode return gives $R_H=D_H/\sqrt{\omega}$; for $\omega\ge0.1$ it agrees with reference-inferred values within $0.21\%$ and tends to the literature high-density limit $1.309283$. Cioslowski and Pernal locate the qualitative crossover to the radially localized strongly correlated harmonium regime at $\omega_{\rm crit}\simeq0.04012$, below the quantitative domain used here. The H$^-$ endpoint and $\omega_{\rm crit}$ are therefore complementary carrier-validity boundaries: they limit the predictive realization, not exact terminal one-mode representability. A broad helium Bethe--Goldstone spectrum confirms that this terminal statement is not microscopic one-pole dynamics.
\end{abstract}

\maketitle

A stationary state is among the most elementary objects in quantum mechanics. If
\begin{equation}
H|\Psi\rangle=E|\Psi\rangle,
\end{equation}
then
\begin{equation}
|\Psi(t)\rangle=e^{-iEt/\hbar}|\Psi(0)\rangle.
\label{eq:stationary}
\end{equation}
The point used here is not the value or observability of an absolute phase, but its multiplicity: a stationary state has one dynamical phase factor, hence one frequency $E/\hbar$ and one scalar energy. Its internal representation may nevertheless be arbitrarily complicated. At the terminal scalar level, a source-supported return with norm $m_0$ and inverse moment $m_{-1}$ has the exact one-mode realization
\[
E_c=-m_{-1}=-\frac{m_0}{D_{\rm eff}},\qquad D_{\rm eff}=\frac{m_0}{m_{-1}}.
\]
This is an exact representation of one scalar energy, not a claim that the microscopic response has one pole. The predictive problem is instead whether the source and its effective terminal mode can be obtained without using the target energy.

For prediction the information carrier must be specified. In the present reg-HF realization, ``mean field'' means the stationary canonical RHF determinant: it has already organized the one-body environment and fixes the occupied representation used to construct the bare source, but it is not asserted to approximate the detailed correlated wave function. Away from exact occupied degeneracies this supplies a reproducible canonical gauge; the exact terminal identity above does not depend on this particular carrier. Primitive Gaussian products in this RHF reference, with exponents $p$ and $q$, define the relative-coordinate scale
\begin{equation}
\rho=\frac{pq}{p+q}
\end{equation}
and the normalized primitive-product relative-coordinate measure
\begin{equation}
P_{\mathbf R}(\mathbf r)=\left(\frac{\rho}{\pi}\right)^{3/2}
\exp[-\rho|\mathbf r-\mathbf R|^2],\qquad
\Sigma=(2\rho)^{-1}I_3.
\label{eq:pairmeasure}
\end{equation}
At fixed covariance this same Gaussian is the common extremizer of several information measures: it minimizes Fisher information through the Fisher--Cram\'er--Rao bound, maximizes Shannon differential entropy, and saturates the three-dimensional Stam inequality,
\begin{equation}
J[P_{\mathbf R}]=\Sigma^{-1},\qquad
S[P]\le S[P_{\mathbf R}],\qquad
N[P_{\mathbf R}]I_F[P_{\mathbf R}]=3.
\label{eq:information}
\end{equation}
These are not independent assumptions imposed on the pair; they identify the same minimum-information Gaussian mode fixed by the HF representation~\cite{Fisher1925,Shannon1948,Stam1959,PanasI}.

Independently, the Coulomb Green function belongs to the same Gaussian scale family. The exact split
\begin{equation}
\frac1r=\frac{\operatorname{erf}(\chi r)}{r}+\frac{\operatorname{erfc}(\chi r)}{r}
\end{equation}
has logarithmic scale derivative
\begin{equation}
K_G(r)=-\chi\frac{\partial}{\partial\chi}\frac{\operatorname{erfc}(\chi r)}{r}
=\frac{2\chi}{\sqrt{\pi}}e^{-\chi^2r^2}.
\label{eq:KG}
\end{equation}
Matching the projected screened-Coulomb and Gaussian channels at coalescence gives
\begin{equation}
\chi^2=a\rho,\qquad a=\frac{\sqrt5-1}{2}.
\label{eq:goldenscale}
\end{equation}
Thus the Gaussian HF primitive-product representation supplies the measure while the Coulomb operator supplies a compatible scale-resolved response.

Regularization alone does not make a stationary correlation energy. The displaced pair response must carry the compensating kinetic response required by scale stationarity. In the retained Gaussian channel the virial supplies
\begin{equation}
T_{\mathrm{vir}}=-\frac12\langle\mathbf r\!\cdot\!\nabla K_G\rangle
=\chi^2\langle r^2K_G\rangle,
\label{eq:virial}
\end{equation}
with restoring envelope $r^2e^{-\chi^2r^2}$. In the dimensionless coordinate $u=\chi r$ this envelope has its unique nonzero maximum at
\begin{equation}
u_\star=1.
\label{eq:unit_shell}
\end{equation}
We therefore regard the Gaussian/Coulomb construction after virial completion as the stationary-compatible pair response selected by the canonical RHF reg-HF realization~\cite{PanasIV}. Up to this point no dimensional reduction of the unrestricted response space has been made.

The exact scalar compression written above already establishes the existence of a one-mode terminal representation. Reg-HF must do more: it must predict that mode independently of the target energy. It does so through the one-mode exclusion rule~\cite{PanasI}
\begin{equation}
P_1:\qquad \Cov_{\perp,R}(u,K_G)\longrightarrow0.
\label{eq:axiom}
\end{equation}
Equation~\eqref{eq:axiom} does not state that orthogonal response components are absent. It specifies the reg-HF representative of the exact terminal mode: covariance carried by orthogonal components is not propagated into the assigned scalar energy or restored by feedback. Thus exact one-mode \emph{existence} and the reg-HF \emph{prediction} of that mode are distinct statements. Feshbach--L\"owdin partitioning provides the familiar precedent that a large complementary space can feed a small effective return~\cite{Feshbach1958,Lowdin1962}; no microscopic one-pole spectrum is implied.

Helium provides an unusually clean test of this predictive realization. There is one occupied spatial pair, no ambiguity in partitioning several occupied pairs, no molecular topology, and no possibility of cancellation among distinct occupied-pair sectors. At the same time, an unlimited set of virtual pair states contributes to the Coulomb ladder. Helium therefore separates \emph{spectral complexity} from \emph{terminal energetic sufficiency}.

After Gaussian averaging, the potential-plus-virial construction gives the dimensionless residual
\begin{equation}
\delta=\frac{7\sqrt5-15}{4}=0.1631189606\ldots .
\label{eq:delta}
\end{equation}
The sequence
\begin{equation}
\rho\longrightarrow\chi^2=a\rho\longrightarrow\text{virial completion}\longrightarrow\delta
\end{equation}
contains no fitted molecular or atomic correlation energy.

Equation~\eqref{eq:axiom} now makes the exact terminal representation predictive. Once orthogonal components are excluded from the reg-HF terminal property, only one scalar Coulombic carrier remains. We write this retained contribution as
\begin{equation}
\Delta_0^{(1)}=-\frac{Z_\star}{r_\star}.
\end{equation}
Neither $Z_\star$ nor $r_\star$ is intended as a literal full-space expectation value. They represent the strength and radial denominator of the retained scalar response. The minimal closure is
\begin{equation}
Z_\star\leftrightarrow\delta,\qquad r_\star\leftrightarrow e,
\end{equation}
so that
\begin{equation}
\boxed{\Delta_0^{(1)}=-\frac{\delta}{e}.}
\label{eq:carrier}
\end{equation}
The appearance of $e$ follows from the unique restoring shell. In the dimensionless coordinate $u=\chi r$, the shell envelope is $u^2e^{-u^2}$, whose nonzero maximum occurs at $u=1$. The Gaussian survival weight at that shell is therefore
\begin{equation}
c_0=e^{-u_\star^2}=e^{-1}.
\end{equation}
The corresponding radial denominator is $r_\star=c_0^{-1}=e$. This identification is a closure hypothesis within the one-mode theory, not a theorem following from the cusp condition alone.

The Coulombic nucleus supplies an additional and essential ingredient. For a He-like ion,
\begin{equation}
H=\sum_{i=1}^2\left(-\frac12\nabla_i^2-\frac{Z}{r_i}\right)+\frac1{r_{12}}.
\end{equation}
Unlike a smooth harmonic trap, the common nucleus introduces singular electron--nucleus channels and therefore participates directly in the return of the displaced pair response. Kato's cusp condition~\cite{Kato1957} supplies the corresponding local Coulombic unit scale. The one-mode closure identifies that Coulombic unit with the unique Gaussian--virial unit shell of Eq.~\eqref{eq:unit_shell}. The absence of an analogous $e^{-1}$ factor in Hooke's atom~\cite{Kestner1962,Taut1993,ONeillGill2003,GillONeill2005} provides an important negative control: the Euler factor is not a generic consequence of Gaussian algebra or of one-mode reduction alone.

The independently known He-like high-density coefficient gives~\cite{Baker1990}
\begin{equation}
c_0^{\mathrm{He}}=0.3678921153\ldots,
\end{equation}
whereas the one-mode closure gives
\begin{equation}
e^{-1}=0.3678794412\ldots .
\end{equation}
The relative difference is $3.45\times10^{-5}$. The value $e^{-1}$ is not chosen from the benchmark; the benchmark is an external test of the independently defined unit-shell closure.

Finite nuclear charge provides a second, independent test. After the L\"owdin radial scaling~\cite{Lowdin1959}
\begin{equation}
\eta=Z-\frac5{16},
\end{equation}
the He-like Hamiltonian can be written as
\begin{equation}
H=\eta^2\left(H_0+\frac1\eta W\right),\qquad \langle0|W|0\rangle=0.
\end{equation}
The uniform first-order Coulomb dilation has therefore been removed, leaving a residual $Q$-space source. Within the projected-source closure derived in Ref.~\cite{PanasIV}, the unit shell has the mean-normalized dilation $q_\star=2/3$; one factor acts on launch and one on return, while either electron can couple to the nucleus. The resulting dimensionless first return insertion is
\begin{equation}
\boxed{\kappa=2q_\star^2\delta=\frac{8\delta}{9}.}
\label{eq:kappa}
\end{equation}
The independently known He-like ratio is
\begin{equation}
\frac{E_3}{|\Delta_0|}\simeq0.1449593,
\end{equation}
whereas Eq.~\eqref{eq:kappa} gives
\begin{equation}
\frac{8\delta}{9}\simeq0.1449946.
\end{equation}
Thus the first finite-$Z$ response tests the retained return stiffness, not merely the asymptotic constant.

Repeated propagation of the same retained mode gives the adopted reg-HF realization of the terminal mode
\begin{equation}
\Delta_{1m}(Z)=-\frac{\delta/e}{1+\kappa/Z},\qquad \kappa=\frac{8\delta}{9}.
\label{eq:onepole}
\end{equation}
Its expansion is
\begin{equation}
\Delta_{1m}(Z)=-\frac{\delta}{e}+\frac{\delta\kappa}{eZ}-\frac{\delta\kappa^2}{eZ^2}+\cdots .
\end{equation}
The important point is conceptual: \emph{rank one does not mean first order}. Equation~\eqref{eq:axiom} excludes propagation through orthogonal response modes in the terminal property, while the retained mode may be propagated repeatedly to all orders. Equation~\eqref{eq:onepole} is therefore a specific predictive realization of the exact one-mode terminal representation, not the definition of one-mode existence itself.

Equation~\eqref{eq:onepole} is implemented without feeding the terminal correction back into the HF determinant. For a finite radial basis $B$, define the stationary HF residual relative to the L\"owdin scale by
\begin{equation}
\Delta_{\rm HF}^{B}(Z)=E_{\rm HF}^{B}(Z)+\left(Z-\frac5{16}\right)^2,
\label{eq:deltaHFbasis}
\end{equation}
and form the conventional one-mode correlation increment
\begin{equation}
E_c^{1m,B}(Z)=\Delta_{1m}(Z)-\Delta_{\rm HF}^{B}(Z).
\label{eq:Ecimplementation}
\end{equation}
The independently contracted bare pair source remains $M_0^{B}=I_{\rm bare}(1,1)$. Equations~\eqref{eq:onepole} and~\eqref{eq:Ecimplementation} therefore determine the terminal pair-return scale through
\begin{equation}
D_{CC}^{1m,B}=\frac{M_0^{B}}{-E_c^{1m,B}},\qquad
E_c^{1m,B}=-\frac{M_0^{B}}{D_{CC}^{1m,B}}.
\label{eq:implementationD}
\end{equation}
Thus Eq.~\eqref{eq:onepole} fixes the return while the converged HF determinant and $M_0^{B}$ remain unchanged. In the reg-HF source convention $M_0^{B}$ is energy-valued, so $D_{CC}$ is a dimensionless confinement factor, not an energy denominator. It is not a second independently fitted observable; it is the effective terminal return implied by the independently specified source and the one-mode energy. We therefore report
\[
R_{CC}(Z)\equiv \frac{D_{CC}(Z)}{Z},
\]
which compares that dimensionless return with the leading nuclear confinement scale.

We evaluated the sequence in element-specific uncontracted HGBS-9 $s$ shells~\cite{Lehtola2020}; H$^-$ uses six additional diffuse $s$ functions obtained by geometric continuation of the two most diffuse official H exponents, with no energy target in the extension. Table~\ref{tab:sequence} gives both the HF basis audit and the one-mode result. The final Gaussian HF energies differ from the independent Laguerre HF values of King, Baskerville, and Cox~\cite{King2018} by less than $0.84\,\mu E_{\mathrm h}$ throughout the sequence, so the correlation comparison is not limited by the radial HF basis. The production $s$-shell counts from H$^-$ through Ar$^{16+}$ are 29, 25, 27, 28, 29, 30, and 32, respectively. For H$^-$, the deterministic diffuse tail is numerically saturated: extending diffuse3 to diffuse4 changes $E_{\rm HF}$ by only $1.35\times10^{-10}E_{\mathrm h}$, with subsequent additions negligible at the displayed precision.

\begin{table*}[t]
\caption{Two-electron Coulomb sequence. H$^-$ uses HGBS-9 plus six target-free diffuse $s$ continuations; the cations use the official element-specific HGBS-9 $s$ subsets. $\Delta E_{\rm HF}=E_{\rm HF}^{B}-E_{\rm HF}^{\rm ref}$ uses the independent Laguerre HF values of Ref.~\cite{King2018}. $D_{CC}$ is the dimensionless terminal pair-return factor of Eq.~\eqref{eq:implementationD}; $R_{CC}=D_{CC}/Z$ compares it with nuclear confinement. The final column is model minus reference.}
\label{tab:sequence}
\centering
\scriptsize
\setlength{\tabcolsep}{3.4pt}
\begin{tabular}{lrrrrrrr}
\hline\hline
Ion & $E_{\rm HF}^{B}/E_{\rm h}$ & $\Delta E_{\rm HF}$ ($\mu E_{\rm h}$) & $E_c^{\rm ref}/E_{\rm h}$ & $E_c^{1m,B}/E_{\rm h}$ & $D_{CC}$ & $R_{CC}$ & error (m$E_{\rm h}$) \\
\hline
H$^-$      & -0.487929733510   & +0.0008 & -0.0398212822 & -0.0371355937 & 0.868588 & 0.868588 & +2.6857 \\
He         & -2.861679989769   & +0.0058 & -0.0420443814 & -0.0419280199 & 1.995352 & 0.997676 & +0.1164 \\
Li$^+$     & -7.236415185856   & +0.0156 & -0.0434982112 & -0.0434826042 & 3.098036 & 1.032679 & +0.0156 \\
Be$^{2+}$  & -13.611299399720  & +0.0309 & -0.0442668078 & -0.0442658392 & 4.195481 & 1.048870 & +0.0010 \\
C$^{4+}$   & -32.361192794463  & +0.0813 & -0.0450537262 & -0.0450556422 & 6.385298 & 1.064216 & -0.0019 \\
Ne$^{8+}$  & -93.861113275933  & +0.2433 & -0.0456929957 & -0.0456934362 & 10.758933 & 1.075893 & -0.0004 \\
Ar$^{16+}$ & -312.861062419263 & +0.8375 & -0.0461228196 & -0.0461224247 & 19.500766 & 1.083376 & +0.0004 \\
\hline\hline
\end{tabular}
\end{table*}

The $D_{CC}$ and $R_{CC}$ columns add a confinement diagnostic to the energy test. Since $M_0$ and $E_c$ carry the same energy unit,
\[
R_{CC}(Z)=\frac{D_{CC}(Z)}{Z}=\frac{M_0(Z)}{Z|E_c(Z)|}
\]
is intrinsically dimensionless. At low $Z$ the Coulomb atom is intrinsically three-body in the two electron--nucleus coordinates and $r_{12}$. With $\mathbf{x}_i=Z\mathbf{r}_i$,
\begin{equation}
\frac{H}{Z^2}=\sum_i\left(-\frac12\nabla_{x_i}^2-\frac1{x_i}\right)+\frac1Z\frac1{x_{12}},
\label{eq:scaledH}
\end{equation}
so increasing $Z$ transfers the dominant nuclear organization to the stationary mean field while the beyond-mean-field residual becomes increasingly pair-relative. Correspondingly, $R_{CC}$ rises from $0.8686$ for H$^-$ to $1.0834$ for Ar$^{16+}$. In atomic units $M_0/Z\to5\delta/16$ while $E_c(\infty)\to-0.0466632540$, giving
\[
R_{CC}(\infty)=\frac{5\delta/16}{0.0466632540}=1.092394\ldots .
\]
The energy unit cancels identically. Thus $Z$ sets the leading confinement scale, but the terminal pair return remains dressed~\cite{Baker1990}. This is structurally analogous to harmonium, where center-of-mass and relative motion separate exactly~\cite{Kestner1962,Taut1993,ONeillGill2003}; it does not imply that the full Coulomb Hamiltonian becomes $r_{12}$-only.

The low-$Z$ endpoint deliberately tests the canonical RHF information carrier rather than the existence of a terminal one-mode representation. King, Baskerville, and Cox find the closed-shell HF critical charge $Z_c^{\rm HF}=1.031177528$~\cite{King2018}, so H$^-$ at $Z=1$ lies outside the binding domain of the doubly occupied HF determinant. Their subsequent pair-density study~\cite{Baskerville2019} and Burton's analysis of the small-$Z$ solutions~\cite{Burton2021} show why this is a genuinely difficult regime: inner/outer radial organization and strong/static correlation are not faithfully represented by one doubly occupied spatial orbital. RHF is nevertheless retained as the symmetry-faithful mean-field information carrier, rather than introducing correlation-like stabilization through a broken-symmetry UHF zeroth-order state. In light of that known complexity, it is noteworthy that the \emph{unchanged} RHF-based reg-HF construction binds H$^-$: $E_{\rm HF}=-0.487929733510\,E_{\mathrm h}$ becomes $E_{\rm regHF}=-0.525065327230\,E_{\mathrm h}$, compared with $E_{\rm exact}\simeq-0.52775101654\,E_{\mathrm h}$. Equation~\eqref{eq:Ecimplementation} therefore recovers $93.3\%$ of the conventional H$^-$ correlation energy. Here $D_{CC}=0.8686<1$ means that the terminal return amplifies the deliberately compact bare source; in the energy-valued source convention this is algebraically equivalent to an enhanced correlation-channel charge $q_{\rm corr}^{\rm eff}=q_e/\sqrt{D_{CC}}$, not to a change of the physical electron charge. The remaining $+2.686$ m$E_{\mathrm h}$ residual is consistent with a regime in which the RHF information carrier is overcompact relative to the diffuse correlated state; it is not evidence against exact terminal one-mode realizability. Conversely, exact agreement from this restricted carrier would not by itself be a cleaner validation, because compensation by the terminal correction could conceal missing reference structure.

The Coulomb sequence and harmonium now provide two deliberately different realizations of the same terminal statement. They share the electron--electron Coulomb singularity, but the external confinement is singular and nuclear in one case and smooth and harmonic in the other. The useful question is therefore not whether their microscopic wavefunctions look alike, but whether a stationary carrier plus one terminal return remains predictive until the carrier ansatz itself becomes qualitatively inadequate.

A change of the \emph{form} of confinement provides the harmonic control. For two electrons in a harmonic trap,
\begin{equation}
H_H(\omega)=\sum_{i=1}^2\left(-\frac12\nabla_i^2+\frac12\omega^2r_i^2\right)+\frac1{r_{12}},
\end{equation}
center-of-mass and relative motion separate exactly~\cite{Kestner1962,Taut1993}. Coordinate scaling shows that the leading interaction parameter is $\omega^{-1/2}$, so $\sqrt{\omega}$ is the harmonic counterpart of $Z$ for comparing return scales. An independent Gaussian-propagation calculation of the complete first-order harmonic response gives
\begin{equation}
\Delta_H^{1m}(\omega)=-\frac{A_H}{1+\kappa_H/\eta_H},\qquad
\omega^2=\eta_H^3\left(\eta_H+\frac{c}{3}\right),
\label{eq:hookepole}
\end{equation}
with $c=\sqrt{2/\pi}$, $A_H=0.0513651488$, and $\kappa_H=0.1847385175$. These coefficients are obtained from the complete first-order oscillator response, not from a finite-$\omega$ energy fit. The harmonic pole is therefore a response-compression control, not the same low-information shell deduction as the Coulomb reg-HF closure. The corresponding conventional increment is formed by the same residual subtraction as in Eq.~\eqref{eq:Ecimplementation}. Published high-accuracy harmonium calculations provide an independent physical boundary and benchmark. Cioslowski and Pernal identify a crossover at $\omega_{\rm crit}\simeq4.011624\times10^{-2}$: for larger $\omega$ the one-electron density remains centered at the trap origin, whereas below it the ground state becomes a radially localized, angularly correlated Wigner-molecule-like pair~\cite{CioslowskiPernal2000}. Matito, Cioslowski, and Vyboishchikov subsequently supplied few-$\mu E_{\mathrm h}$ FCI benchmarks over the useful confinement range~\cite{Matito2010}. We therefore restrict the quantitative one-mode comparison to $\omega\ge0.1$, conservatively inside the centered weak-correlation/strong-confinement regime.

To place this successful harmonic one-mode return on the \emph{same reg-HF source--return interface} as the Coulomb sequence, we retain the independently defined co-centered source
\begin{equation}
M_0^H(\omega)=\frac{\delta}{2}J_{ii}^{\rm HF}(\omega)
\end{equation}
and define the equivalent dimensionless return factors
\begin{equation}
D_H^{1m}=\frac{M_0^H}{-E_c^{1m,H}},\qquad
R_H^{1m}(\omega)=\frac{D_H^{1m}}{\sqrt{\omega}}.
\label{eq:hookeR}
\end{equation}
Using the same source with an independent reference correlation energy gives $D_H^{\rm ref}$ and $R_H^{\rm ref}$. This translation is a moment-interface representation, not a derivation of the harmonic denominator from $\delta$ alone. The reference energies are exact Taut checkpoints where available and otherwise independent radial values consistent with the Cioslowski--Pernal and Matito--Cioslowski--Vyboishchikov benchmarks~\cite{Taut1993,CioslowskiPernal2000,Matito2010}. Table~\ref{tab:hookeR} shows that the independently constructed one-mode return reproduces the reference-inferred confinement ratio closely throughout moderate and strong confinement.

\begin{table}[t]
\caption{Hooke-atom terminal return on the reg-HF source interface. $R_H=D_H/\sqrt{\omega}$. The reference column combines the same stationary HF source with exact Taut or independent numerical total energies consistent with Refs.~\cite{CioslowskiPernal2000,Matito2010}; $R_H^{1m}$ uses Eq.~\eqref{eq:hookepole}. The finite-$\omega$ reference ratio is diagnostic, not an input to the one-mode construction. The last column is $100(R_H^{1m}/R_H^{\rm ref}-1)$.}
\label{tab:hookeR}
\centering
\footnotesize
\setlength{\tabcolsep}{4.5pt}
\begin{tabular}{rrrr}
\hline\hline
$\omega$ & $R_H^{\rm ref}$ & $R_H^{1m}$ & rel. error (\%) \\
\hline
0.10 & 1.863238 & 1.867024 & +0.203 \\
0.50 & 1.545092 & 1.543371 & -0.111 \\
1.00 & 1.473040 & 1.471784 & -0.085 \\
10   & 1.359145 & 1.358944 & -0.015 \\
100  & 1.324824 & 1.324801 & -0.002 \\
$\infty$ & 1.309283 & 1.309283 & 0 \\
\hline\hline
\end{tabular}
\end{table}

The high-density endpoint follows without a finite-$\omega$ fit. Since $J_{ii}^{\rm HF}\sim\sqrt{2\omega/\pi}$,
\begin{equation}
R_H(\infty)=\frac{\delta/\sqrt{2\pi}}{|E_c^H(\infty)|}=1.3092825\ldots,
\label{eq:hookeinf}
\end{equation}
using the independently established harmonic high-density correlation limit $E_c^H(\infty)=-0.04970283317E_{\rm h}$~\cite{LoosGill2009}. The Coulomb and harmonic limits, $R_{CC}(\infty)=1.092394\ldots$ and $R_H(\infty)=1.309283\ldots$, therefore show that nearly similar high-density correlation energies need not arise from the same terminal return scale: the effective one-mode denominator retains the physics of the confining field. The weak-trap limit is deliberately excluded. As $\omega$ approaches and crosses the Cioslowski--Pernal boundary, the simple scale-stationary product carrier no longer contains the qualitative radial localization of the exact state, and Eq.~\eqref{eq:hookepole} correspondingly fails to reproduce the correct leading $\omega^{2/3}$ behavior. H$^-$ is the Coulombic cousin of this $\omega_{\rm crit}$ boundary: not the same physical transition, but the same diagnostic event for the predictive theory---a qualitative reorganization not represented by the chosen mean-field information carrier. This does not challenge the exact scalar identity $E_c=-m_0/D_{\rm eff}$; it limits the predictive realization of $M_0$ and $D_{\rm eff}$ from that carrier. Cioslowski's partial-wave analysis independently shows that the microscopic angular structure becomes progressively more demanding as confinement weakens~\cite{Cioslowski2015}.

The remaining question is whether ``one mode'' has merely been confused with an accidentally one-pole microscopic response. Bethe--Goldstone (BG) theory~\cite{Bethe1956,Goldstone1957,Nesbet1967,Freed1968} provides the crucial control.

For a fixed starting energy, let the Hermitian positive BG return matrix be $A_{\rm BG}$ and its source vector $v$. Define
\begin{equation}
m_0=v^\dagger v,\qquad
m_{-1}=v^\dagger A_{\rm BG}^{-1}v,\qquad
m_{-2}=v^\dagger A_{\rm BG}^{-2}v.
\label{eq:bgmoments}
\end{equation}
The fully resummed ladder energy is then exactly
\begin{equation}
E_c^{\rm BG}=-m_{-1}=-\frac{m_0}{D_{\rm BG}},\qquad
D_{\rm BG}=\frac{m_0}{m_{-1}}.
\label{eq:bgcompression}
\end{equation}
Equivalently, after diagonalizing $A_{\rm BG}$, $D_{\rm BG}^{-1}=\langle D^{-1}\rangle_p$ is the source-weighted mean inverse microscopic denominator. Equation~\eqref{eq:bgmoments} is therefore not the same source convention as Eq.~\eqref{eq:implementationD}: $m_0=v^\dagger v$ has energy-squared units, whereas the reg-HF $M_0$ is energy-valued. The two are nevertheless the same terminal moment architecture in different source normalizations. For any energy scale $\Lambda$ define
\[
M_0^{(\Lambda)}=\frac{m_0}{\Lambda},\qquad
D^{(\Lambda)}=\frac{D_{\rm BG}}{\Lambda},\qquad
m_{-1}=\frac{M_0^{(\Lambda)}}{D^{(\Lambda)}}.
\]
Taking $\Lambda=m_0/M_0^B$ makes the normalized BG source equal to the independently constructed reg-HF source and produces a dimensionless BG return. This is a representation conversion, not a derivation of the reg-HF denominator: Eq.~\eqref{eq:implementationD} uses its separately predicted inverse moment $-E_c^{1m,B}$. Thus the full ladder is exactly one-mode at the scalar-energy level without being one-pole microscopically.

To distinguish scalar compression from literal one-pole dynamics, define
\begin{equation}
W_{\mathrm{BG}}=\frac{m_0m_{-2}}{m_{-1}^2}-1.
\label{eq:W}
\end{equation}
With source weights $p_\nu=|\langle\nu|v\rangle|^2/m_0$, this has the exact form
\begin{equation}
W_{\mathrm{BG}}=
\frac{\operatorname{Var}_{p}(D^{-1})}{\langle D^{-1}\rangle_{p}^{2}},
\label{eq:Wvariance}
\end{equation}
so $W_{\mathrm{BG}}$ is the squared coefficient of variation of the inverse return denominator. For a positive return measure, $W_{\mathrm{BG}}=0$ only when the source sees a single denominator, apart from degeneracy. An explicit helium BG calculation gives
\begin{equation}
W_{\mathrm{BG}}\simeq1.03\text{--}1.06,
\end{equation}
depending on radial discretization. The archived $18s12p10d8f$ virtual basis is even tempered, $\alpha_n=\alpha_0\beta^{n-1}$, with starting exponents $0.020$, $0.020$, $0.030$, and $0.040$ bohr$^{-2}$ for $s,p,d,f$. Thus $\beta$ is only a basis-discretization ratio, not a physical coupling or fit parameter. Over $\beta=2.05$--$2.25$, $E_c^{\rm BG}$ changes from $-0.04191825$ to $-0.04193428E_{\rm h}$, $D_{\rm BG}$ from $10.26996$ to $10.44105E_{\rm h}$, and $W_{\rm BG}$ from $1.02656$ to $1.05937$. The broad-width conclusion is therefore stable.

The two systems therefore supply a paired test. In the Coulomb series, He and the positive ions combine a broad microscopic return, exact scalar one-mode compression, and an independently constructed reg-HF mode with high energy accuracy; H$^-$ then exposes the boundary of the canonical RHF information carrier while the unchanged construction still recovers $93.3\%$ of the correlation energy. In harmonium, the response-compressed one-mode return is comparably accurate for $\omega\ge0.1$, while the Cioslowski--Pernal weak-confinement crossover identifies the regime in which the scale-stationary product carrier ceases to represent the qualitative state. The common pattern is therefore not ``one mode works only for weakly correlated systems.'' Rather, exact one-mode terminal representability survives, while a particular predictive realization becomes limited when its chosen stationary information carrier is no longer qualitatively adequate.

The universal statement concerns the terminal scalar realization, not a one-dimensional microscopic response. The Coulomb values $\delta/e$ and $8\delta/9$ belong to the present predictive closure. In the present reg-HF implementation, the mean-field information carrier is the canonical RHF determinant; that representation choice constructs $M_0$ without claiming that RHF reproduces the detailed correlated wave function. The resulting proposition is:
\par\smallskip
\noindent\setlength{\fboxsep}{4pt}%
\fbox{\parbox{\dimexpr\columnwidth-2\fboxsep-2\fboxrule\relax}{\centering\footnotesize\bfseries
Exact terminal one-mode representability;\\
predictive realization is carrier- and closure-dependent.}}
\par\smallskip
The two controls make the limitation concrete. H$^-$ is the Coulombic cousin of the harmonium $\omega_{\rm crit}$ boundary: in the former, the symmetry-faithful RHF information carrier is too compact to represent the diffuse inner/outer organization; in the latter, the centered product carrier fails as the pair becomes radially localized. The processes are physically different, but both mark failure of a chosen carrier before any failure of the exact terminal identity. Thus the exact statement is $E_c=-m_0/D_{\rm eff}$, whereas reg-HF predicts an energy-valued $M_0$ and its return from a specified stationary information carrier.

\begin{acknowledgments}
OpenAI ChatGPT assisted with editing and document preparation; the author is responsible for all scientific content and verification.
\end{acknowledgments}

\end{document}